\documentstyle[aps,prl,epsfig,psfig]{revtex}
\tightenlines
   \newcommand{\be}{\begin{equation}}
   \newcommand{\ee}{\end{equation}}
   \newcommand{\bea}{\begin{eqnarray}}
   \newcommand{\eea}{\end{eqnarray}}

\def\init{\tabskip 0pt\offinterlineskip}
\def\crr{\cr\noalign{\hrule}}

\begin{document}

\draft
\widetext
\title{Dynamic Spin Correlation Function near the Antiferromagnetic Quantum Phase 
Transition of Heavy-Fermions}
\author{C. P\'epin$^1$ and M. Lavagna$^2$}
\address{Commissariat \`a l'Energie Atomique, \\ D\'epartement de
  Recherche Fondamentale sur la Mati\`ere Condens\'ee/SPSMS, \\
17, rue des Martyrs,
  \\ 38054 Grenoble Cedex 9, France}

\maketitle \widetext
  \leftskip 10.8pt
  \rightskip 10.8pt
  \begin{abstract}
 The dynamical spin susceptibility is studied in the magnetically-disordered phase of heavy-Fermion systems near the antiferromagnetic quantum phase transition. In the framework of the $S=1/2$ Kondo lattice model, we introduce a perturbative expansion treating the spin and Kondo-like degrees of freedom on an equal footing. The dynamical spin susceptibility displays a two-component behaviour in agreement with the inelastic neutron scattering (INS) experiments performed in $CeCu_{6}$, $Ce_{1-x}La_{x}Ru_{2}Si_{2}$ and $UPt_{3}$: a quasielastic q-independent peak as in a Fermi liquid theory, and a strongly q-dependent inelastic peak typical of a non-Fermi
liquid behaviour. Very strikingly, the position of the inelastic peak is found to be pushed to
zero at the antiferromagnetic transition. The picture is consistent with the neutron cross sections observed in INS experiments.\par
  \end{abstract}

\bigskip 

\section{\protect\bigskip Introduction}

One of the most striking properties of heavy Fermion compounds discovered
these last years is the existence of a quantum phase transition \cite{lohneysen,kambe} driven by
composition change (at $x_{C}=0.1$ in $CeCu_{6-x}Au_{x}$ and $x_{C}=0.08$ in
$Ce_{1-x}La_{x}Ru_{2}Si_{2}$), pressure or magnetic field. It has been largely discussed in various theoretical approaches \cite{moriya,rosch}. Important
insight is provided by the evolution of the low temperature neutron cross
section measured by Inelastic Neutron Scattering (INS) experiments when
getting closer to the magnetic instability. The experiments performed
in pure compounds $CeCu_{6}$ and $CeRu_{2}Si_{2}$ by Regnault et al \cite{regnault} and Aeppli et al \cite{aeppli} have shown the presence of two distinct contributions to
the dynamic magnetic structure factor: a $\bf{q}$-independent
quasielastic component, and a strongly $\bf{q}$-dependent
inelastic peak with a maximum at the value $\omega _{\max }$ of the
frequency. The former corresponds to localized excitations of Kondo-type
while the latter peaked at some wavevector ${\bf{Q}}$ is
believed to be associated with intersite magnetic correlations due to RKKY
interactions. The frequency-width of the quasielastic and inelastic peaks
respectively define the single-site and intersite relaxation rate $\Gamma
_{SS}$ and $\Gamma _{IS}$. Such features have also been observed in $UPt_{3}$
and called as ''slow'' and ''fast'' components by Bernhoeft and Lonzarich \cite{bernhoeft}.
Later on, INS\ experiments have been performed with varying compositions as
in $Ce_{1-x}La_{x}Ru_{2}Si_{2}$ \cite{raymond}. It has been observed a narrowing of
the single-site relaxation rate $\Gamma _{SS}$ when getting closer to the
magnetic transition. At the same time, both the position of
the inelatic peak $\omega _{\max }$ and the intersite relaxation rate $%
\Gamma _{IS}$ drastically decrease when getting near the magnetic
instability. Table 1 reports the values of $\Gamma _{SS}$, $\Gamma _{IS}$
and $\omega _{\max }$ for the different compounds.

Any theory aimed to describe the quantum critical phenomena in heavy-Fermion
compounds should account for the so-quoted behaviour of the dynamical spin
susceptibility. We start from the Kondo lattice model which is believed to
describe the physics of these systems. We refer to the recent paper of
Tsunetsugu et al \cite{tsunetsugu} for a review of the model. As already pointed out by
Doniach in his initial paper \cite{doniach77}, the main features result of the competition
between the Kondo effect and the RKKY interactions among spins mediated by
the conduction electrons. Most of the theories developed so far 
\cite{lacroix79,brandow,fazekas,rice,goltsev} agree with
the existence of a hybridization gap which splits the Abrikosov-Suhl or
Kondo resonance formed at the Fermi level. The role of the interband
transitions has been outlined for long in order to explain the inelatic
component of the dynamical spin susceptibility. For instance, the theories
based on a 1/N expansion \cite{coleman,read,millis,auerbach86,lavagna} (where N is simultaneously the degeneracy of the
conduction electrons and of the spin channels) predict a maximum of $\chi
^{"}(k_{F},\omega )/{\omega}$ at $\omega _{\max }$ of the order of the indirect
hybridization gap \cite{auerbach88}. However, the 1/N expansion theories present serious
drawbacks: (i) the spin fluctuation effects are automatically ruled out
since the RKKY interactions only occur at the following order in $1/N^{2}$ \cite{houghton},
(ii) they then fail to describe any magnetic instability and hence the
quantum critical phenomena mentioned above and (iii) the predictions for $%
\omega _{\max }$ and the associated relaxation rate cannot account for the
experimental observations near the magnetic instability. An improvement
brought by Doniach \cite{doniach87} consists to consider the $%
1/N^{2}$ corrections in an instantaneous approximation: it gives back the
ladder diagram contribution to the dynamical spin susceptibility and then accounts
for the spin fluctuation effects. Other approaches were proposed in Ref. \cite{lacroix91,evans1,evans2}. But still the
predictions for the frequency dependence of the dynamic magnetic structure factor
presents a gap of the order of the hybridization gap whatever the value of
the interaction is. On the other hand, in front of the difficulties
encountered when starting from microscopic descriptions, various
phenomenological models (as the duality model of Kuramoto and Miyake \cite{kuramoto90} 
and reference \cite{bernhoeft}) have
been introduced to describe both the spin fluctuation and the
itinerant electron aspects with some successful predictions as the weak
antiferromagnetism of these systems.

In this paper, we develop a systematic approach to the Kondo lattice model
for $S=1/2$ ($N=2$) in which the Kondo-like and the spin degrees of freedom
are treated on an equal footing. The presented approach shows some similarities
with earlier works \cite{lacroix79,goltsev}. But while Ref.\cite{lacroix79,goltsev}
essentially describe the phase diagram of the Kondo lattice at a mean-field level,
we focus on the effects of spin fluctuations in the magnetically-disordered
phase hence bringing the spin-fluctuation and the Kondo effect theories 
together. The saddle-point results and the gaussian
fluctuations in the charge channel are consistent with the standard $1/N$
theories. In addition, the gaussian fluctuations in the spin channel restore
the spin fluctuation effects which were missing in the $1/N$
expansion. The general expression of the dynamical spin
susceptibility that we derive reproduces some of the features postulated in
the phenomenological models. It presents a two-component behaviour: a
quasielatic component superimposed on an inelatic peak with renormalized
values of the relaxation rates, susceptibilities and $\omega _{\max }$. In a
very striking way, $\omega _{\max }$ is pushed to zero and the inelastic
mode becomes soft at the antiferromagnetic phase transition with vanishing
relaxation rate. Predictions are quantitatively compared with
experimental results. The quasielastic peak is typical of a Fermi liquid
while the other mode breaks the Fermi liquid description. Our approach might
offer new prospects for the study of the quantum critical phenomena in the 
vicinity of the antiferromagnetic phase transition.

\bigskip 

\section{\protect\bigskip Presentation of the approach}

We consider the Kondo lattice model (KLM) constituted by a periodic array of
Kondo impurities with an average number of conduction electrons per site $%
n_{c}<1$. In the grand canonical ensemble, the hamiltonian is
written as

\begin{equation}
\label{eq1}
H=\sum_{k\sigma }\varepsilon _{k}c_{k\sigma }^{\dagger}c_{k\sigma
}+J\sum\limits_{i}{\bf{S}}_{i} \cdot \sum\limits_{\sigma \sigma ^{\prime
}}c_{i\sigma }^{\dagger}\mbox{\boldmath{$\tau$}}_{\sigma \sigma'} c_{i\sigma ^{\prime } }-\mu N_S(\frac{1}{N_{S}}%
\sum_{k\sigma }c_{k\sigma }^{\dagger}c_{k\sigma }-n_{c})
\end{equation}

in which $\bf{S}_{i}$ represents the spin ($S=1/2$) of the
impurities distributed on the sites (in number $N_{S}$) of a
periodic lattice; $c_{k\sigma }^{\dagger}$ is the creation operator of the conduction
electron of momentum $\bf{k}$, spin quantum number $\sigma $
characterized by the energy $\epsilon _{k}=-\sum\limits_{<i,j>}t_{ij}\exp \left( i%
{\bf{k}}.{\bf{R}}_{ij}\right) $ and the chemical
potential $\mu $; $\mbox{\boldmath{$\tau$}}$ are the Pauli matrices $%
\left(\mbox{\boldmath{$\tau$}} ^{x}, \mbox{\boldmath{$\tau$}} ^{y}, \mbox{\boldmath{$\tau$}} ^{z}\right) $ and $\mbox{\boldmath{$\tau$}}^{0}$ the unit matrix; J is the antiferromagnetic
Kondo interaction $\left( J>0\right) $.

We use the Abrikosov pseudo-fermion representation of the spin $%
{\bf{S}}_{i}$: ${\bf{S}}_{i}=\sum\limits_{\sigma \sigma
^{\prime }}f_{i\sigma }^{\dagger}\mbox{\boldmath{$\tau$}}_{\sigma \sigma ^{\prime
}}f_{i\sigma ^{\prime }}$. The projection into the physical subspace is
implemented by a local constraint  

\be
\label{eq2}
Q_{i}=\frac{1}{N_{S}}%
\sum\limits_{i\sigma }f_{i\sigma }^{+}f_{i\sigma }-1=0
\ee

A Lagrange multiplier $\lambda _{i}$ is introduced to enforce the local
constraint $Q_{i}$. Since $[Q_{i},H]=0$, $\lambda _{i}$ is time-independent.

In this representation, the partition function of the KLM can be expressed
as a functional integral over the coherent states of the fermion fields 

\begin{equation}
\label{eq3}
Z=\int {\cal D}c_{i\sigma }{\cal D}f_{i\sigma }d\lambda _{i}\exp \left[-\int_{0}^{\beta }{\cal L}(\tau
)d\tau\right]  
\end{equation}

where the Lagrangian ${\cal L}(\tau)$ is given by

\[
{\cal L}(\tau )={\cal L}_{0}(\tau )+H_{0}(\tau )+H_{J}(\tau ) 
\]

\[
{\cal L}_{0}(\tau )=\sum_{i\sigma }c_{i\sigma }^{\dagger}\partial _{\tau }c_{i\sigma
}+f_{i\sigma }^{\dagger}\partial _{\tau }f_{i\sigma }
\]

\[
H_{0}(\tau )=\sum_{k\sigma }\epsilon _{k}c_{k\sigma }^{\dagger}c_{k\sigma }-\mu
N_{S}\left( \frac{1}{N_{S}}\sum_{k\sigma }c_{k\sigma }^{\dagger}c_{k\sigma
}-n_{c}\right) +\sum_{i}\lambda _{i}Q_i 
\]

\[
H_{J}(\tau )=J\sum_{i}{\bf{S}}_{fi} \cdot {\bf{S}}_{ci}
\]
with ${\bf{S}}_{c_{i}}=\sum\limits_{\sigma \sigma ^{\prime
}}c_{i\sigma }^{\dagger}\mbox{\boldmath{$\tau$}}_{\sigma \sigma ^{\prime
}}c_{i\sigma ^{\prime }}$ and ${\bf{S}}_{fi}={\bf{S}}_{i}$

\par
We perform a Hubbard-Stratonovich transformation on the Kondo
interaction term $H_{J}(\tau )$. Since more than one field is implied in the transformation, 
an uncertainty is left on the way of decoupling. We propose to remove it in the following way. 
First, we note that $H_{J}(\tau)$ may also be written as 

\be
\label{eq4}
H_{J}(\tau )=-\frac{3J}{8}\sum_{i}n_{fc_{i}}n_{cf_{i}}+\frac{J}{2}\sum_{i}%
{\bf{S}}_{fc_{i}} \cdot {\bf{S}}_{cf_{i}}
\ee

where ${\bf{S}}_{fc_{i}}=\sum\limits_{\sigma \sigma ^{\prime
}}f_{i\sigma }^{\dagger}\mbox{\boldmath{$\tau$}}_{\sigma \sigma ^{\prime
}}c_{i\sigma ^{\prime }}$ and $n_{fc_{i}}=\sum\limits_{\sigma \sigma ^{\prime }}f_{i\sigma }^{\dagger} {\tau}_{\sigma \sigma ^{\prime}}^{0} c_{i\sigma ^{\prime}}$
(respectively ${\bf{S}}_{cf_{i}}$ and $n_{cf_{i}}$ their hermitian conjugate).

\bigskip 

The Kondo interaction term is then given by any linear combination of 
$J\sum\limits_{i}{\bf{S}}_{fi}\cdot {\bf{S}}_{ci}$ (with a
weighting factor x) and of the term appearing in the right-hand side of
Equation (\ref{eq4}) (with a weighting factor (1-x)). x is chosen so as to recover the
usual results obtained within the slave-boson theories. One can check that this is
the case for $x=1/3$.
The Kondo interaction term is then given by 

\be
\label{eq5}
H_{J}(\tau )=J_{S}\sum_{i}\left( {\bf{S}}_{f_{i}}\cdot {\bf{%
S}}_{c_{i}}+{\bf{S}}_{fc_{i}}\cdot {\bf{S}}%
_{cf_{i}}\right) -J_{C}\sum_{i}n_{fc_{i}}n_{cf_{i}}
\label{cinq}
\ee

with $J_{S}=J/4$ and $J_{C}=J/3$.

\bigskip
Performing a generalized Hubbard-Stratonovich transformation on the
partition function Z makes the fields $\Phi _{i}$, $\Phi
_{i}^{\ast }$ (for charge) and $\mbox{\boldmath{$\xi$}}_{f_{i}},$ $%
\mbox{\boldmath{$\xi$}}_{c_{i}}$ appear (omitting the fields associated to $%
{\bf{S}}_{fc_{i}},$ ${\bf{S}}_{cf_{i}})$. We get

\be
\label{eq6}
Z=\int d\Phi _{i}d\Phi _{i}^{\ast }d\mbox{\boldmath{$\xi$}}_{f_{i}}d%
\mbox{\boldmath{$\xi$}}_{c_{i}}{\cal D}c_{i\sigma }{\cal D}f_{i\sigma }d\lambda _{i}\exp \left[-\int_{0}^{\beta }{\cal L}^{\prime}(\tau
)d\tau\right]  
\ee
with 
\[
{\cal L}^{\prime}(\tau )={\cal L}_{0}(\tau )+H_{0}(\tau )+H^{\prime}_{J}(\tau ) 
\]

\[
H^{\prime }_{J}(\tau )=\sum_{i\sigma \sigma ^{\prime }}
\left( 
\begin{array}{cc}
c_{i\sigma }^{\dagger} & f_{i\sigma }^{\dagger}
\end{array}
\right) 
\left( 
\begin{array}{cc}
-J_{S}i\mbox{\boldmath{$\xi$}}_{f_{i}}\cdot \mbox{\boldmath{$\tau$}}_{\sigma \sigma ^{\prime
}} & J_{C}\Phi
_{i}^{\ast}\tau^{0}_{\sigma \sigma ^{\prime
}} \\ 
J_{C}\Phi _{i}\tau^{0}_{\sigma \sigma ^{\prime
}} & -J_{S}i\mbox{\boldmath{$\xi$}}_{c_{i}}\cdot 
\mbox{\boldmath{$\tau$}}_{\sigma \sigma ^{\prime
}}
\end{array}
\right) 
\left( 
\begin{array}{c}
c_{i\sigma ^{\prime }} \\ 
f_{i\sigma ^{\prime }}
\end{array}
\right)  +J_{C}\sum_{i}\Phi _{i}^{\ast }\Phi _{i}+J_{S}\sum_{i}%
\mbox{\boldmath{$\xi$}}_{f_{i}}.\mbox{\boldmath{$\xi$}}_{c_{i}}
\]

\subsection{\protect\bigskip Saddle-Point}

The saddle-point solution is obtained for space and time independent fields $%
\Phi _{0\text{ }}$, $\lambda _{0}$, $\xi _{f_{0}}$ and $\xi _{c_{0}}$. 
In the magnetically-disordered regime 
($\xi _{f_{0}}=\xi _{c_{0}}=0)$, it leads to renormalized bands $\alpha 
$ and $\beta $ as schematized in Figure 1. Noting  $\sigma _{0}^{(\ast)}=J_C \Phi _{0}^{(\ast)}$ and
$\varepsilon _{f}=\lambda _{0}$, $\alpha _{k\sigma }^{\dagger}|0>$ and $%
\beta _{k\sigma }^{\dagger}|0>$ are the eigenstates of

\begin{equation}
\label{eq7}
{\bf G}_{0}^{-1 \sigma }(\bf{k},\tau )=\left( 
\begin{array}{cc}
\partial _{\tau }+\varepsilon _{k} & \sigma _{0}^{\ast} \\ 
\sigma _{0} & \partial _{\tau }+\varepsilon _{f}
\end{array}
\right) 
\end{equation}
with respectively the eigenenergies $\left( \partial _{\tau }+E_{k}^{-}\right)$ 
and $\left( \partial _{\tau }+E_{k}^{+}\right) $. In the notations: 
$x_{k}=\varepsilon _{k}-\varepsilon _{f}$, 
$y_{k}^{\pm }=E_{k}^{\pm }-\varepsilon _{f}$ and $\Delta _{k}=\sqrt{x_{k}^{2}+4\sigma _{0}^{2}}$, we get

\begin{equation}
y_{k}^{\pm }=\left( x_{k}\pm \Delta _{k}\right) /2
\end{equation}
Let us note $U_{k\sigma }^{\dagger}$ the matrix transforming the initial basis $%
(c_{k\sigma }^{\dagger}$ $f_{k\sigma }^{\dagger})$ to the eigenbasis $(\alpha _{k\sigma }^{\dagger}$ $%
\beta _{k\sigma }^{\dagger})$. The hamiltonian being hermitian, the matrix $U_{k\sigma
}$ is unitary : $U_{k\sigma }U_{k\sigma }^{\dagger}=U_{k\sigma }^{\dagger}U_{k\sigma }=1$. In
the notation $U_{k\sigma }^{\dagger}=\left( 
\begin{array}{cc}
-v_{k} & u_{k} \\ 
u_{k} & v_{k}
\end{array}
\right) $, we have 
\be
\label{eq9}
u_{k}=\frac{-\sigma _{0}/y_{k}^{-}}{\sqrt{1+(\sigma _{0}/y_{k}^{-})^{2}}}=%
\frac{1}{2}\left[ 1+\frac{x_{k}}{\Delta _{k}}\right] 
\ee
\[
v_{k}=\frac{1}{\sqrt{1+(\sigma _{0}/y_{k}^{-})^{2}}}=\frac{1}{2}\left[ 1-%
\frac{x_{k}}{\Delta _{k}}\right] 
\]
The saddle-point equations together with the conservation of the number of
conduction electrons are written as 
\begin{equation}
\label{eq10}
\sigma _{0}=\frac{1}{N_{S}}J_{C}\sum_{k\sigma }u_{k}v_{k}~n_{F}(E_{k}^{-})
\end{equation}
\[
1=\frac{1}{N_{S}}\sum_{k\sigma }u_{k}^{2}~n_{F}(E_{k}^{-}) 
\]
\[
n_{c}=\frac{1}{N_{S}}\sum_{k\sigma }v_{k}^{2}~n_{F}(E_{k}^{-}) 
\]
Their resolution leads to
\be
\label{eq11}
\left| y_{F}\right| =D\exp \left[ -2/\left( \rho _{0}J_C\right) \right] 
\ee
\[
2\rho _{0}\sigma _{0}^{2}/\left| y_{F}\right| =1 
\]
\[
\mu =0 
\]
where $y_{F}=\mu -\varepsilon _{F}$ and $\rho _{0}$ is the bare density of
states of conduction electrons ($\rho _{0}=1/2D$ for a flat band). Noting 
$y=E-\varepsilon _{F}$, the density of states at the energy E is $\rho \left( E\right) =\rho _{0}\left( 1+\sigma _{0}^{2}/y^{2}\right)$. If $n_c<1$, the chemical potential is located just below the upper edge 
of the $\alpha$-band. The system is metallic. The density of states at the Fermi level 
is strongly enhanced towards the bare density of states of conduction electrons : 
$\rho (E_{F})/\rho_{0}=(1+\sigma _{0}^{2}/y_{F}^{2})\sim 1/(2\rho _{0}\left| y_{F}\right| )$.
That corresponds to the flat part of the $\alpha$-band in Figure 1. It is
associated to the formation of a Kondo or Abrikosov-Suhl resonance pinned at the Fermi level resulting of the Kondo effect. The low-lying excitations are 
quasiparticles of large effective mass $m^{\ast }$ as observed in heavy-Fermion
systems. Also note the presence of a
hybridization gap between the $\alpha $ and the $\beta $ bands. The direct
gap of value $2\sigma _{0}$ is much larger than the indirect gap equal to 2$%
\left| y_{F}\right| $. The saddle-point solution transposes to N=2 the
large-N results obtained within the slave-boson mean-field theories (SBMFT).

\subsection{\protect\bigskip Gaussian fluctuations}

We now consider the gaussian fluctuations around the saddle-point solution.
Following Read and Newns \cite{read}, we take advantage of the local U(1) gauge
transformation of the lagrangian ${\cal L}^{\prime}(\tau)$

\[
\Phi _{i}\rightarrow r_{i}\exp (i\theta _{i}) 
\]
\[
f_{i}\rightarrow f_{i}^{\prime }\exp (i\theta _{i}) 
\]
\[
\lambda _{i}\rightarrow \lambda _{i}^{^{\prime }}+i~\partial \theta
_{i}/\partial \tau 
\]
We use the radial gauge in which the modulus of both fields $\Phi _{i}$ and $%
\Phi _{i}^{\ast }$ are the radial field $r_{i}$, and their phase $\theta
_{i} $ (via its time derivative) is incorporated into the Lagrange
multiplier $\lambda _{i}$ which turns out to be a field. Use of the radial
instead of the cartesian gauge bypasses the familiar complications of
infrared divergences associated with unphysical Goldstone bosons. We let the
fields fluctuate away from their saddle-point values : $r_{i}=r_{0}+\delta
r_{i}$, $\lambda _{i}=\lambda _{0}+\delta \lambda _{i}$, $\bf{%
\xi }_{f_{i}}=\delta \mbox{\boldmath{$\xi$}}_{f_{i}}$ and $\bf{\xi 
}_{c_{i}}=\delta \mbox{\boldmath{$\xi$}}_{c_{i}}$. After integrating out the
Grassmann variables in the partition function in Equation (\ref{eq6}), we get 

\begin{equation}
\label{eq12}
Z=\int {\cal D}r_{i}{\cal D} \lambda _{i}{\cal D}\mbox{\boldmath{$\xi$}}_{f_{i}}{\cal D}\mbox{\boldmath{$\xi$}}_{c_{i}}\exp
[-S_{eff}]
\end{equation}
where the effective action is

\[
S_{eff}=-\sum_{k,i\omega _{n}}Ln~Det{\bf G}^{-1}({\bf{k}},i\omega
_{n})+\beta ~[~J_{C}\sum_{i}r_{i}^{2}+J_{S}\sum_{i}\mbox{\boldmath{$\xi$}}%
_{f_{i}} \cdot \mbox{\boldmath{$\xi$}}_{c_{i}}+N_{S}(\mu n_{c}-\lambda _{0})]
\]
with : 
\[
\left[ {\bf G}^{-1}(i\omega _{n})\right] _{ij}^{\sigma \sigma ^{\prime }}=\left( 
\begin{array}{cc}
\lbrack (-i\omega _{n}-\mu )\delta _{ij}-t_{ij}]\delta _{\sigma \sigma
^{\prime }}-J_{S}i\mbox{\boldmath{$\xi$}}_{f_{i}}.\mbox{\boldmath{$\tau$}}%
_{\sigma \sigma ^{\prime }}\delta _{ij} & (\sigma _{0}+J_{C}\delta
r_{i})\delta _{\sigma \sigma ^{\prime }}\delta _{ij} \\ 
(\sigma _{0}+J_{C}\delta r_{i})\delta _{\sigma \sigma ^{\prime }}\delta _{ij}
& [-i\omega _{n}+\varepsilon _{f}+\delta \lambda _{i}]\delta _{\sigma \sigma
^{\prime }}\delta _{ij}-J_{S}i\mbox{\boldmath{$\xi$}}_{c_{i}}.\mbox{\boldmath
{$\tau$}}_{\sigma \sigma ^{\prime }}\delta _{ij}
\end{array}
\right) 
\]
Expanding up to the second order in the Bose fields, one obtains the
gaussian corrections $S_{eff}^{(2)}$ to the saddle-point effective action 
\bea
\label{eq13}
S_{eff}^{(2)} &=&\frac{1}{\beta} \sum_{{\bf{q}},i\omega_{\nu}}[\left( 
\begin{array}{cc}
\delta r & \delta \lambda 
\end{array}
\right) {\bf D}_{C}^{-1}({\bf{q}},i\omega_{\nu})\left( 
\begin{array}{c}
\delta r \\ 
\delta \lambda 
\end{array}
\right)  +\left( 
\begin{array}{cc}
\delta \xi _{f}^{z} & \delta \xi _{c}^{z}
\end{array}
\right) {\bf D}_{S}^{\Vert -1}({\bf{q}},i\omega_{\nu})\left( 
\begin{array}{c}
\delta \xi _{f}^{z} \\ 
\delta \xi _{c}^{z}
\end{array}
\right)  \nonumber \\
&+&\left( 
\begin{array}{cc}
\delta \xi _{f}^{+} & \delta \xi _{c}^{+}
\end{array}
\right) {\bf D}_{S}^{\bot -1}({\bf{q}},i\omega_{\nu})\left( 
\begin{array}{c}
\delta \xi _{f}^{-} \\ 
\delta \xi _{c}^{-}
\end{array}
\right) +\left( 
\begin{array}{cc}
\delta \xi _{f}^{-} & \delta \xi _{c}^{-}
\end{array}
\right) {\bf D}_{S}^{\bot -1}({\bf{q}},i\omega_{\nu})\left( 
\begin{array}{c}
\delta \xi _{f}^{+} \\ 
\delta \xi _{c}^{+}
\end{array}
\right)] 
\eea
where the boson propagators split into the following charge and longitudinal spin parts 
\be
\label{eq14}
{\bf D}_{C}^{-1}({\bf{q}},i\omega_{\nu})=\left( 
\begin{array}{cc}
J_{C}[1-J_{C}(\overline{\varphi }_{2}({\bf{q}},i\omega_{\nu})+\overline{\varphi }_{m}({\bf{q}},i\omega_{\nu}))] & 
-J_{C}\overline{\varphi }_{1}({\bf{q}},i\omega_{\nu}) \\ 
-J_{C}\overline{\varphi }_{1}({\bf{q}},i\omega_{\nu}) & -\overline{\varphi }_{ff}({\bf{q}},i\omega_{\nu})
\end{array}
\right) 
\ee
\[
{\bf D}_{S}^{\Vert -1}({\bf{q}},i\omega_{\nu})=\left( 
\begin{array}{cc}
J_{S}^{2}\varphi _{ff}^{\Vert }({\bf{q}},i\omega_{\nu}) & J_{S}[1+J_{S}\varphi _{cf}^{\Vert }({\bf{q}},i\omega_{\nu})]
\\ 
J_{S}[1+J_{S}\varphi _{fc}^{\Vert }({\bf{q}},i\omega_{\nu})] & J_{S}^{2}\varphi _{cc}^{\Vert }({\bf{q}},i\omega_{\nu})
\end{array}
\right) 
\]
and equivalent expression for the transverse spin part ${\bf D}_{S}^{\bot -1}({\bf{q}},i\omega_{\nu})$. The
expression of the different bubbles are given in the appendix. The charge boson propagator
${\bf D}_{C} ({\bf{q}},i\omega_{\nu})$ associated to the Kondo effect is equivalent to that obtained in
the $1/N$ expansion theories. The originality of the approach is to simultaneously derive
the spin propagator ${\bf D}_{S}^{\Vert -1}({\bf{q}},i\omega_{\nu})$ and ${\bf D}_{S}^{\bot -1}({\bf{q}},i\omega_{\nu})$ associated to the spin fluctuation effects.
Note that in the magnetically-disordered phase, the charge and spin contributions in $S_{eff}$
are totally decoupled. 

\section{\protect\bigskip Dynamical spin susceptibility}

Next step is to consider the dynamical spin susceptibility. For that
purpose, we study the linear response $M_{f}$ to the source-term $-2%
{\bf{S}}_{f}.\bf{B}$ (we consider $\bf{B}$
colinear to the $\bf{z}$-axis). The effect on the partition
function expressed in Equation (\ref{eq6}) is to change the hamiltonian $H^{\prime }_{J}(\tau )$ to $%
H^{\prime B}_{J}(\tau )$  
\be
\label{eq15}
H^{\prime B}_{J}(\tau )=\sum_{i\sigma \sigma ^{\prime
}}\left( 
\begin{array}{cc}
c_{i\sigma }^{\dagger} & f_{i\sigma }^{\dagger}
\end{array}
\right) \left( 
\begin{array}{cc}
-J_{S}i\mbox{\boldmath{$\xi$}}_{f_{i}}\cdot {\mbox{\boldmath{$\tau$}}}_{\sigma \sigma ^{\prime
}} & J_{C}\Phi
_{i}^{\ast }\tau^{0}_{\sigma \sigma ^{\prime
}} \\ 
J_{C}\Phi _{i}\tau^{0}_{\sigma \sigma ^{\prime
}} & \sum\limits_{\alpha =x,y,z}(-J_{S}i\xi
_{c_{i}}^{\alpha }-B\delta _{\alpha z}).\tau^{\alpha}_{\sigma \sigma ^{\prime
}}
\end{array}
\right) \left( 
\begin{array}{c}
c_{i\sigma ^{\prime }} \\ 
f_{i\sigma ^{\prime }}
\end{array}
\right) +  J_{C}\sum_{i}\Phi _{i}^{\ast }\Phi _{i}+J_{S}\sum_{i}%
\mbox{\boldmath{$\xi$}}_{f_{i}}.\mbox{\boldmath{$\xi$}}_{c_{i}}
\ee
Introducing the change of variables $\xi _{c_{i}}^{\alpha }=\xi
_{c_{i}}^{\alpha }-iB/J_{S}$, we connect the f magnetization and the ff
dynamical spin susceptibility to the Hubbard Stratonovich fields $%
\mbox{\boldmath{$\xi$}}_{f_{i}}$ 
\[
M_{f}^{z}=-\frac{1}{\beta }\frac{\partial LnZ}{\partial B_{z}}=i\left\langle
\xi _{f_{i}}^{z}\right\rangle 
\]
\begin{equation}
\label{eq16}
\chi _{ff}^{\alpha \beta }=-\frac{1}{\beta }\frac{\partial ^{2}LnZ}{\partial
B^{\alpha }\partial B^{\beta }}=-
\left\langle \xi _{f_{i}}^{\alpha} \xi _{f_{i}}^{\beta}\right\rangle + 
\left\langle  \xi _{f_{i}}^{\alpha} \right\rangle  \left\langle \xi _{f_{i}}^{\beta}\right\rangle 
\end{equation}
Using the expression (\ref{eq14}) fot the boson propagator ${\bf D}_{S}^{\Vert -1}({\bf q})$, we get for the longitudinal spin susceptibility
\begin{equation}
\label{eq17}
\chi _{ff}^{\Vert }({\bf{q}},i\omega_{\nu})=\frac{\varphi _{ff}^{\Vert }({\bf{q}},i\omega_{\nu})}{1-J_{S}^{2}[\varphi
_{ff}^{\Vert }({\bf{q}},i\omega_{\nu})\varphi _{cc}^{\Vert }({\bf{q}},i\omega_{\nu})-\varphi _{fc}^{\Vert 2}({\bf{q}},i\omega_{\nu})-\frac{2}{%
J_{S}}\varphi _{fc}^{\Vert }({\bf{q}},i\omega_{\nu})]}
\end{equation}
and equivalent expression for the transverse spin susceptibility $\chi
_{ff}^{\bot }({\bf{q}},i\omega_{\nu})$. The diagrammatic representation of Equation (\ref{eq17}) is reported in Figure 2. The different bubbles $\varphi _{ff} ({\bf{q}},i\omega_{\nu})$, $\varphi _{cc} ({\bf{q}},i\omega_{\nu})$
and $\varphi _{fc} ({\bf{q}},i\omega_{\nu})$ are evaluated from the expressions of the Green's
functions 
\be
\label{eq18}
G_{ff}({\bf{k}},i\omega_{n})=u_{k}^{2}G_{\alpha \alpha }({\bf{k}},i\omega_{n})+v_{k}^{2}G_{\beta
\beta }({\bf{k}},i\omega_{n})
\ee
\[
G_{cc}({\bf{k}},i\omega_{n})=v_{k}^{2}G_{\alpha \alpha }({\bf{k}},i\omega_{n})+u_{k}^{2}G_{\beta
\beta }({\bf{k}},i\omega_{n})
\]
\[
G_{cf}({\bf{k}},i\omega_{n})=G_{fc}({\bf{k}},i\omega_{n})=-u_{k}v_{k}[G_{\alpha \alpha}
({\bf{k}},i\omega_{n})-G_{\beta \beta }({\bf{k}},i\omega_{n})]
\]
where $G_{\alpha \alpha }({\bf k},i\omega_n)$ and $G_{\beta \beta }({\bf k},i\omega_n)$
are the Green's functions associated to the eigenstates $\alpha _{k\sigma }^{\dagger}|0>$%
and $\beta _{k\sigma }^{\dagger}|0>$. In the low frequency limit, one can easily check that the
dynamical spin susceptibility may be written as

\begin{equation}
\label{eq19}
\chi _{ff}({\bf{q}},i\omega_{\nu})=\frac{\chi _{\alpha \alpha }({\bf{q}},i\omega_{\nu})+\overline{\chi}
_{\alpha \beta }({\bf{q}},i\omega_{\nu})}{1-J_{S}^{2}\chi _{\alpha \alpha }({\bf{q}},i\omega_{\nu})%
\overline{\chi} _{\alpha \beta }({\bf{q}},i\omega_{\nu})}
\end{equation}
for both the longitudinal and the transverse parts. 
\[
\chi _{\alpha \alpha }({\bf{q}},i\omega_{\nu})=\frac{1}{\beta }\sum\limits_{k}\frac{%
n_{F}(E_{k}^{-})-n_{F}(E_{k+q}^{-})}{i\omega _{\nu }-E_{k+q}^{-}+E_{k}^{-}}
\]
\[
\overline{\chi} _{\alpha \beta }({\bf{q}},i\omega_{\nu})=\frac{1}{\beta }\sum%
\limits_{k}(u_{k}^{2}v_{k+q}^{2}+v_{k}^{2}u_{k+q}^{2})\frac{%
n_{F}(E_{k}^{-})-n_{F}(E_{k+q}^{+})}{i\omega _{\nu }-E_{k+q}^{+}+E_{k}^{-}}
\]
Equation (\ref{eq19}) constitutes the main result of the paper from which the whole physical
discussion on the $\bf{q}$- and $\omega $- dependence of the
dynamical spin susceptibility follows and comparison with experiments is
made.

\section{\protect\bigskip Physical discussion}
 
From Equation (\ref{eq19}), one can see that the dynamical spin susceptibility is made
of two contributions $\chi _{intra}({\bf{q}},i\omega_{\nu})$ and $\chi _{inter}({\bf{q}},i\omega_{\nu})$

\be
\label{eq20}
\chi _{ff}({\bf{q}},i\omega_{\nu})=\chi _{intra}({\bf{q}},i\omega_{\nu})+\chi _{inter}({\bf{q}},i\omega_{\nu}) 
\ee

with

\be
\label{eq21}
\chi _{intra}({\bf{q}},i\omega_{\nu})=\frac{\chi _{_{\alpha \alpha }}({\bf{q}},i\omega_{\nu})}{1-J_{S}^{2}\chi
_{_{\alpha \alpha }}({\bf{q}},i\omega_{\nu})\overline{\chi }_{\alpha \beta }({\bf{q}},i\omega_{\nu})}
\ee

\be
\label{eq22}
\chi _{inter}({\bf{q}},i\omega_{\nu})=\frac{\overline{\chi}_{\alpha \beta }({\bf{q}},i\omega_{\nu})}{%
1-J_{S}^{2}\chi _{_{\alpha \alpha }}({\bf{q}},i\omega_{\nu})\overline{\chi }_{\alpha \beta }({\bf{q}},i\omega_{\nu})}
\ee

$\chi _{intra}({\bf{q}},i\omega_{\nu})$and $\chi _{inter}({\bf{q}},i\omega_{\nu})$ respectively represent the
renormalized particle-hole pair excitations within the lower $\alpha $ band, and
from the lower $\alpha $ to the upper $\beta $ band. The latter expression is
reminiscent of the behaviour proposed by Bernhoeft and Lonzarich \cite{bernhoeft} to explain
the neutron scattering observed in $UPt_{3}$ with the existence of both a "slow" 
and a "fast" component in $\chi^{"}(\bf{q},\omega)/{\omega}$ due to spin-orbit 
coupling. Also in a phenomenological way, the same type of feature has been suggested 
in the duality model developed by Kuramoto and Miyake \cite{kuramoto90}. To our knowledge, the proposed approach
provides the first microscopic derivation from the Kondo lattice model of
such a behaviour.
The bare intraband susceptibility $\chi _{_{\alpha \alpha }}({\bf{q}},\omega )$ is well approximated by a lorentzian

\be
\label{eq23}
\chi _{_{\alpha \alpha }}^{-1}({\bf{q}},\omega )=\rho _{\alpha \alpha }({\bf{q}})^{-1} \left( 1-i\frac{%
\omega }{\Gamma _{0}({\bf{q}})}\right) 
\ee

where $\rho _{\alpha \alpha }=\chi _{_{\alpha \alpha }}^{\prime}({\bf{q}},0)$ and $\Gamma _{0}(\bf{q})$ is  the relaxation rate of order $\left| y_{F}\right|=T_{K}$. This
corresponds to the Lindhard continuum of the intraband particle-hole pair 
excitations $\chi _{_{\alpha \alpha }}^{"}(q,\omega)\neq 0$ as reported in Figure 3. 
In the same way, we propose to schematize the low-frequency behavior $(\omega<<\omega _{0}({\bf{q}})$ of the bare interband susceptibility by

\be
\label{eq26}
\overline{\chi }_{\alpha \beta }^{\prime,-1}({\bf{q}},\omega )=\rho _{\alpha \beta}({\bf{q}})^{-1}\left( 1-\frac{%
\omega }{\omega _{0}({\bf{q}})}\right) 
\ee

where $\rho _{\alpha \beta}=\overline{\chi }_{_{\alpha \beta}}^{\prime}({\bf{q}},0)$ and $\omega _{0}(\bf{q})$ is a characteristic frequency-scale of the interband transitions. The value of $\omega _{0}(\bf{q})$
is strongly structure-dependent. In the simple case of a cubic band 
structure $\epsilon_{k}=-2t(\cos k_{x}+\cos k_{y}+\cos k_{z})$ 
(tight-binding scheme including nearest-neighbor hopping), we find 
a weakly wavevector dependent frequency around 
${\bf{q}}={\bf{Q}}$ of order of 
$\omega _{0}=2\left| y_{F}\right| /\left( \rho _{0}J_{C}\right)$. 
The latter result does not stand for more complicated band structures
as obtained by de Haas-van Alphen studies \cite {julian} combined with band structure 
calculations in heavy-Fermion compounds. 
In the following, we will leave $\omega _{0}(\bf{q})$ as a parameter.
Figure 3 reports the continuum of interband particle-hole excitations 
$\overline{\chi}_{\alpha \beta }"\neq 0$. Due to the presence 
of the hybridization gap in
the density of states, the latter continuum displays a gap equal to $2\sigma
_{0}$, the value of the direct gap at ${\bf{q}}={\bf{0}}$%
, and $2\left| y_{F}\right| $, the value of the indirect gap at 
${\bf{q}}={\bf{Q}}$ (close to $k_{F}$). More precisely,
we have shown

\be
\label{eq27}
\overline{\chi }_{\alpha \beta }"({\bf{0}},\omega )=4\rho _{0}%
\frac{\sigma _{0}^{2}}{\omega \sqrt{\omega ^{2}-4\sigma _{0}^{2}}}\text{ at }%
2\sigma _{0}<\omega <D
\ee

\[
\overline{\chi }_{\alpha \beta }"({\bf{Q}},\omega )=2\rho _{0}%
\frac{1}{1+\omega ^{2}/(2\sigma _{0})^{2}}\text{ at }2\left| y_{F}\right|
<\omega <2D
\]

Far from the antiferromagnetic wavevector $\bf{Q}=(\pi,\pi,\pi)$, $\chi_{ff} ({\bf{q}},\omega )$
is dominated by the intraband transitions. In the low frequency limit, the frequency dependence of $\chi _{intra}^{"}(\bf{q},\omega)$ can be approximate to a lorentzian 

\be
\label{eq24}
\chi_{ff}^{"} ({\bf{q}},\omega ) \approx \chi _{intra}^{"}({\bf{q}},\omega)=\omega  \frac{\chi
_{intra}^{^{\prime }}({\bf{q}})\Gamma _{intra}({\bf{q}})}{\omega ^{2}+{\Gamma
_{intra}({\bf{q}})}^{2}} 
\ee

with

\be
\label{eq25}
\Gamma _{intra}({\bf{q}})=\Gamma _{0}({\bf{q}})(1-I({\bf{q}})) 
\ee

\[
\chi _{intra}^{^{\prime }}({\bf{q}})=\frac{\rho _{\alpha \alpha }({\bf{q}})}{(1-I({\bf{q}}))} 
\]

$I({\bf{q}})=J_{S}^{2}\chi_{\alpha \alpha}^{\prime}({\bf{q}},0) \overline{\chi }_{\alpha \beta }^{\prime}
({\bf{q}},0)$. One has: $\chi_{\alpha \alpha}^{\prime}({\bf{0}},0)=\rho_{\alpha \alpha }({\bf{0}})=\rho(E_F)$ and $\chi_{\alpha \beta}^{\prime}({\bf{0}},0)=\rho_{0}$.
The contribution expressed in equation (\ref{eq24}) is
consistent with the standard Fermi liquid theory. Note that the product $\Gamma _{intra}({\bf{q}})\chi _{intra}^{^{\prime }}({\bf{q}})=\rho_{\alpha\alpha}({\bf{q}})\Gamma_{0}({\bf{q}})$ is independent
of I. 

\bigskip

Oppositely, at the antiferromagnetic wavevector $\bf {Q}$, $\chi_{ff} ({\bf{q}},\omega )$
is driven by the interband contribution and we get

\be
\label{eq28}
\chi_{ff}^{"} ({\bf{Q}},\omega ) \approx \chi _{inter}^{"}({\bf{Q}},\omega )=\omega \frac{I\chi
_{inter}^{\prime }\Gamma _{inter}}{(\omega -\omega _{\max })^{2}+\Gamma
_{inter}^{2}}
\ee

with

\be
\label{eq29}
\omega _{\max } =\omega _{0}(1-I)
\ee

\[
\Gamma _{inter} =\omega _{0}^{2}(1-I)/\Gamma _{0}
\]

\[
\chi _{inter}^{\prime } =\rho _{\alpha \beta }/(1-I)
\]

where  $\omega _{0}$, $\rho _{\alpha \beta }$, $\Gamma_{0}$ and $I$ are the
values of $\omega _{0}(\bf{q})$, $\rho _{\alpha \beta }(\bf{q})$
and $\Gamma_{0}(\bf{q})$ and $I(\bf{q})$ at $\bf{q}=\bf{Q}$.
The role of the interband transitions have already been pointed out in
previous works \cite{auerbach88}. However while the previous studies conclude to the presence
of an inelastic peak at finite value of the frequency related to the
hybridization gap whatever the interaction J is, we emphasize that the
renormalization of $\overline{\chi }_{\alpha \beta }({\bf{Q}},\omega )$ into $\chi _{inter}
({\bf{Q}},\omega )$ leads to a noteworthy renormalization of
the interband gap.
Due to the damping introduced by intraband transitions, $\chi _{inter}^{"}(\bf{Q},\omega )$
takes a finite value at frequency much smaller than the hybridization gap. Both the relaxation rate $%
\Gamma _{inter}$ vanishes and the susceptibility $\chi _{inter}^{\prime }$
diverges at the antiferromagnetic transition with again the product $\Gamma_{inter} \chi_{inter}^{\prime }$
independent of $I$. Remarkably, the value $%
\omega _{\max }$ of the maximum of $\chi _{inter}^{"}(\bf{Q},\omega )/{\omega}$
is at the same time pushed to zero. Such a behaviour has
been effectively observed in $Ce_{1-x}La_{x}Ru_{2}Si_{2}$ \cite{raymond} with a reduction of $\Gamma _{inter}$
and $\omega _{\max }$ respectively by a factor 4 and 6 when x goes from 0 to 0.075 so when
getting closer to the magnetic instability occuring at $x=0.08$. In order
to make the comparison more quantitative, we propose to deduce the values of
$\omega _{0}$ and $(1-I)$ from the experimental data using the equations (\ref{eq29}): 
$\omega _{0}={\Gamma _{0}\Gamma _{inter}}/{\omega _{\max}}$
and $(1-I)={\omega _{\max}^2}/({\Gamma _{0}\Gamma _{inter}})$.
Table II reports the results starting from  the experimental values of $\Gamma _{0}, \Gamma _{inter}$, $%
\omega _{\max }$ (respectively noted $\Gamma _{SS}, \Gamma _{IS}$, $
\omega _{0}$ in experimental papers) extracted from the INS results 
obtained in $CeCu_6$ (ref. \cite {regnault}) and $Ce_{1-x}La_{x}Ru_2Si_2$ at $x=0$
and $x=0.075$ (ref. \cite {raymond}). The predictions for $\omega _{0}$ and $(1-I)$
in these compounds seem reasonable. The Stoner enhancement factor $(1-I)$
decreases in $Ce_{1-x}La_{x}Ru_2Si_2$ from $x=0$ to $x=0.075$. $(1-I)$ of $CeCu_6$ is intermediate between those two systems.

\bigskip 

\section{\protect\bigskip Conclusion}

In this paper, we have set up a new approach of the $S=1/2$ Kondo lattice
model which enlarges the standard $1/N$ expansion theories up on the spin
fluctuation effects. The latter effects are proved to be essential for the
behaviour of the dynamical spin susceptibility near the magnetic phase
transition. Our approach provides a microscopic derivation of the main
features assumed in the phenomenogical models of heavy Fermions as the
duality model. We predict a two-component behaviour of the dynamical spin
susceptibility: a quasielastic peak typical of the Fermi liquid excitations,
and an inelastic peak at a value $\omega _{\max }$ of the frequency which is
strongly renormalized due to spin fluctuation effects. Outstandingly well, the
frequency of the inelastic peak is pushed to zero at the antiferromagnetic
transition at the same time as the frequency width vanishes. The results
have been compared to the Inelastic Neutron Scattering experiment data with
reasonable predictions for the Stoner enhancement factor $(1-I)$ and the characteristic
frequency  $\omega _{0}$ of the interband contribution to the susceptibility. 
Obviously, more experiments are needed for a systematic test. The issue is important 
since it may have implications for the quantum critical phenomena around the antiferromagnetic
critical point. Work is currently in progress in that direction and will be presented in a forthcoming
paper. We expect the two underlined modes to have different effects on the critical behaviour with,
on the one hand,  the first "intraband" mode acting as a paramagnon mode as in 
the Hertz-Moriya-Millis theory \cite {moriya}, and on the other hand, additional effects brought 
by the second "interband" mode. 

\vspace{0.4in}
\centerline {\bf ACKNOWLEDGEMENTS}
\vspace{0.2in}
We would like to thank G.G. Lonzarich, N.R. Bernhoeft, G.J. McMullan, L.P. Regnault, 
J. Flouquet, S. Raymond, P. Brison and K. Miyake for very helpful discussions.

\vspace{0.4in}
\centerline {\bf APPENDIX}
\vspace{0.2in}

The expressions of the different bubbles appearing in the expression of the
boson propagators (cf. Eq.\ref{eq14}) are given here (with i=1, 2, m or ff)
\be
\overline{\varphi }_{i}({\bf{q}},i\omega_{\nu})=\varphi
_{i}({\bf{q}},i\omega_{\nu})+\varphi _{i}(-{\bf{q}},-i\omega_{\nu})
\ee
\begin{eqnarray*}
\varphi _{1}({\bf{q}},i\omega_{\nu}) &=&-\frac{1}{\beta} \sum_{k\sigma,i\omega_n}G_{cf_{0}}^{\sigma
}({\bf k+q},i\omega_{n}+i\omega_{\nu})G_{ff_{0}}^{\sigma }({\bf{k}},i\omega_{n}) \\
\varphi _{2}({\bf{q}},i\omega_{\nu}) &=&-\frac{1}{\beta} \sum_{k\sigma,i\omega_n}G_{cc_{0}}^{\sigma
}({\bf k+q},i\omega_{n}+i\omega_{\nu})G_{ff_{0}}^{\sigma }({\bf{k}},i\omega_{n}) \\
\varphi _{m}({\bf{q}},i\omega_{\nu}) &=&-\frac{1}{\beta} \sum_{k\sigma,i\omega_n}G_{cf_{0}}^{\sigma
}({\bf k+q},i\omega_{n}+i\omega_{\nu})G_{cf_{0}}^{\sigma }({\bf{k}},i\omega_{n})
\end{eqnarray*}
\begin{eqnarray*}
\varphi _{ff}^{\Vert }({\bf{q}},i\omega_{\nu}) &=&-\frac{1}{\beta } \sum_{k\sigma,i\omega_n}G_{ff_{0}}^{\sigma}({\bf k+q},i\omega_{n}+i\omega_{\nu})G_{ff_{0}}^{\sigma }({\bf{k}},i\omega_{n}) \\
\varphi _{cc}^{\Vert }({\bf{q}},i\omega_{\nu}) &=&-\frac{1}{\beta } \sum_{k\sigma,i\omega_n}G_{cc_{0}}^{\sigma }({\bf k+q},i\omega_{n}+i\omega_{\nu})G_{cc_{0}}^{\sigma }({\bf{k}},i\omega_{n})\\
\varphi _{fc}^{\Vert }({\bf{q}},i\omega_{\nu}) &=&-\frac{1}{\beta } \sum_{k\sigma,i\omega_n}G_{fc_{0}}^{\sigma }({\bf k+q},i\omega_{n}+i\omega_{\nu})G_{fc_{0}}^{\sigma }({\bf{k}},i\omega_{n})
\end{eqnarray*}
\begin{eqnarray*}
\varphi _{ff}^{\bot }({\bf{q}},i\omega_{\nu}) &=&-\frac{1}{\beta } \sum_{k\sigma,i\omega_n}G_{ff_{0}}^{\uparrow }({\bf k+q},i\omega_{n}+i\omega_{\nu})G_{ff_{0}}^{\downarrow }({\bf{k}},i\omega_{n}) \\
\varphi _{cc}^{\bot }({\bf{q}},i\omega_{\nu}) &=&-\frac{1}{\beta } \sum_{k\sigma,i\omega_n}G_{cc_{0}}^{\uparrow }({\bf k+q},i\omega_{n}+i\omega_{\nu})G_{cc_{0}}^{\downarrow }({\bf{k}},i\omega_{n}) \\
\varphi _{fc}^{\bot }({\bf{q}},i\omega_{\nu}) &=&-\frac{1}{\beta } \sum_{k\sigma,i\omega_n}G_{fc_{0}}^{\uparrow }({\bf k+q},i\omega_{n}+i\omega_{\nu})G_{fc_{0}}^{\downarrow }({\bf{k}},i\omega_{n})
\end{eqnarray*}
where $G_{cc_{0}}^{\sigma }({\bf{k}},i\omega_{n})$, $%
G_{ff_{0}}^{\sigma }({\bf{k}},i\omega_{n})$ and $G_{fc_{0}}^{\sigma }({\bf{k}},i\omega_{n})$ are the Green's
functions at the saddle-point level obtained by inversing the matrix $%
G_{0}^{\sigma }(\bf{k},\tau )$ defined in Equation (\ref{eq7}).

\bigskip

$^1$ Present Address: Department of Physics, MIT Ma02139 Cambridge, US
\par
$^2$ Also Part of the Centre National de la Recherche Scientifique (CNRS)

\vfill\eject

\centerline {\bf TABLE CAPTIONS}

\vspace{0.4in}

Table I: Values of the single-site and intersite relaxation rates $\Gamma_{SS}$ and
$\Gamma_{IS}$ and position of the inelastic peak $\omega_{max}$ extracted from
the Inelastic Neutron Scattering (INS) measurements performed in $CeCu_6$ (ref. \cite{regnault})
and $Ce_{1-x}La_{x}Ru_2Si_2$ at $x=0$ and $x=0.075$ (ref. \cite{raymond}).

\vspace{0.2in} 

Table II: Predicted values of the characteristic frequency-scale $\omega_{0}$ for the interband 
transitions and the Stoner enhancement factor $(1-I)$ from the INS data on $\Gamma _{SS}$, $\Gamma _{IS}$ and $\omega _{0}$  (respectively noted $\Gamma_{0}$, $\Gamma _{inter}$ and $\omega _{\max }$ in experimental papers) for the same three compounds as in Table I. Note that $Ce_{1-x}La_{x}Ru_2Si_2$ at $x=0.075$ is very close to the antiferromagnetic instability while the Stoner enhancement factor $(1-I)$ for 
$CeCu_6$ is intermediate between that of the two concentrations $x=0$ and $x=0.075$ of $Ce_{1-x}La_{x}Ru_2Si_2$.

\vfill\eject

\vspace{0.4in}

\centerline {\bf Table I}
 
$$\vbox{\init\halign to 14cm{
\strut#&\vrule#\tabskip=1em plus 2em&
\hfil$#$\hfil&
\vrule$\,$\vrule#&
\hfil$#$\hfil&
\vrule#&
\hfil$#$\hfil&
\vrule#&
\hfil$#$\hfil&
\vrule#\tabskip 0pt\crr
&& &&\Gamma_{SS}(meV)&&\Gamma_{IS}(meV)&&\omega_{max}(meV)\hfil&\crr
&&CeCu_6&&0.42&&0.2&&0.25&\crr
&&CeRu_2Si_2&&2.0&&0.75&&1.2&\crr
&&Ce_{1-x}La_xRu_2Si_2&&1.4&&0.2&&0.2&\crr
}}$$

\vspace {0.4in}

\centerline {\bf Table II}

$$\vbox{\init\halign to 18,3cm{
\strut#&\vrule#\tabskip=1em plus 2em&
\hfil$#$\hfil&
\vrule$\,$\vrule#&
\hfil$#$\hfil&
\vrule#&
\hfil$#$\hfil&
\vrule#&
\hfil$#$\hfil&
\vrule#&
\hfil$#$\hfil&
\vrule#&
\hfil$#$\hfil&
\vrule#\tabskip 0pt\crr
&& &&\Gamma_{0}(meV)&&\Gamma_{inter}(meV)&&\omega_{max}(meV)&&\omega_{0}
|_{ded.}(meV)&&(1-I)|_{ded.}\hfil&\crr
&&CeCu_6&&0.42&&0.2&&0.25&&0.34&&0.74&\crr
&&CeRu_2Si_2&&2.0&&0.75&&1.2&&1.25&&0.96&\crr
&&Ce_{1-x}La_xRu_2Si_2&&1.4&&0.2&&0.2&&1.4&&0.14&\crr
}}$$

\vfill\eject

\centerline {\bf FIGURE CAPTIONS}

\vspace{0.4in}

Figure 1: Energy versus wave-vector k for the two bands $\alpha$ and $\beta$. Note 
the presence of a direct gap of value $2\sigma _{0}$ and of an indirect gap of
value $2\left| y_{F}\right|$.

\vspace{0.2in}

Figure 2: Diagrammatic representation of Equation (\ref{eq27}) for the dynamical spin susceptibility $\chi _{ff}(\bf{q},\omega)$.

\vspace{0.2in}

Figure 3: Continuum of the intra- and interband electron-hole pair excitations 
$\chi _{_{\alpha \alpha}}^{"}(q,\omega)\neq 0$ and $\chi _{_{\alpha \beta}}^{"}(q,\omega)\neq 0$.
Note the presence of a gap in the interband transitions equal to the indirect gap of value 
$2\left| y_{F}\right|$ at $q=k_F$, and to the direct gap of value $2\sigma _{0}$ at $q=0$.
 
\vfill\eject

\vspace{2in}
\begin{figure}
\centerline{\psfig{file=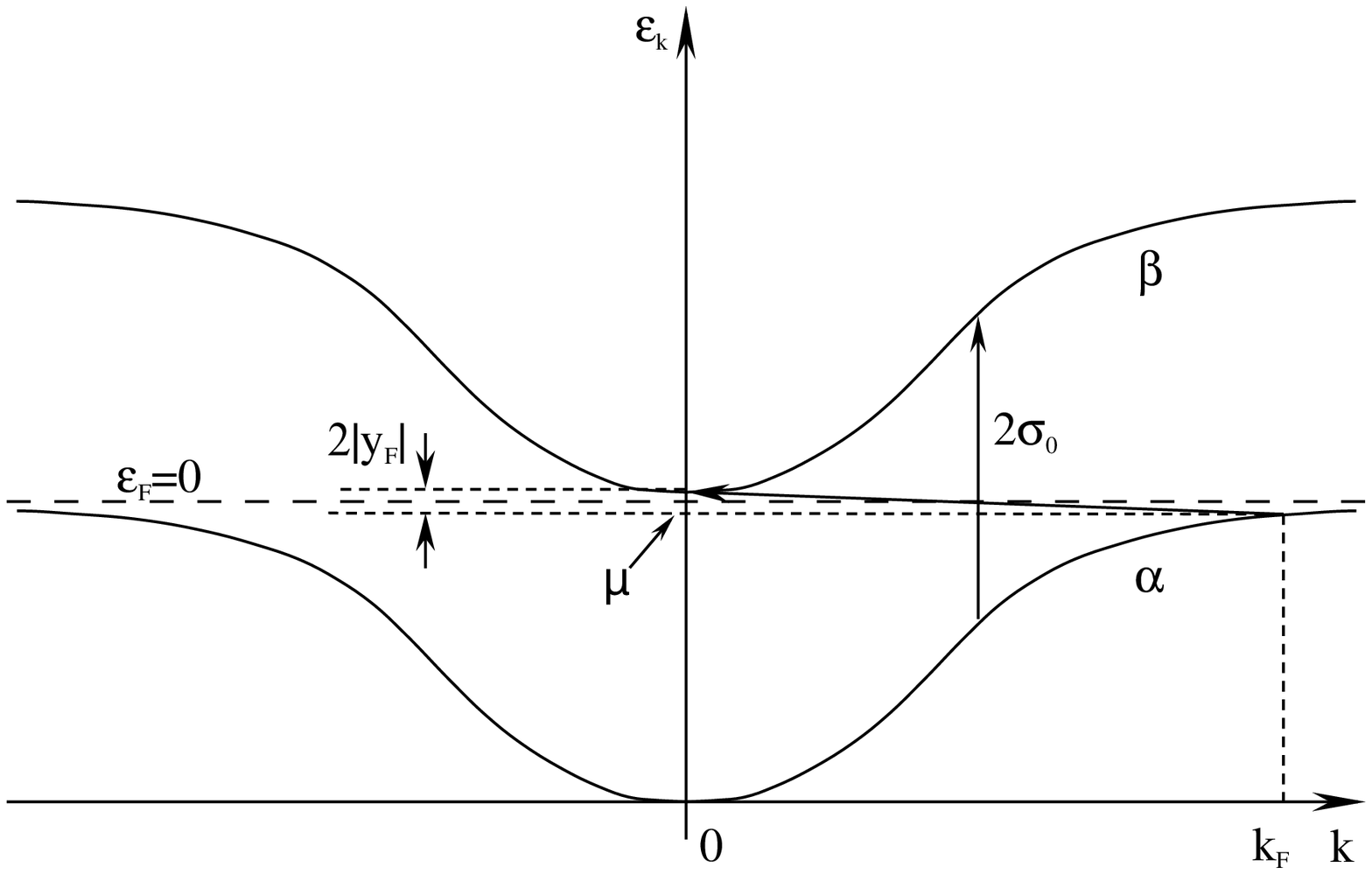,height=9cm,width=16cm}}
\vspace{2in}
\caption{}
\label{fig1}
\end{figure}

\vfill\eject

\vspace{2in}
\begin{figure}
\centerline{\psfig{file=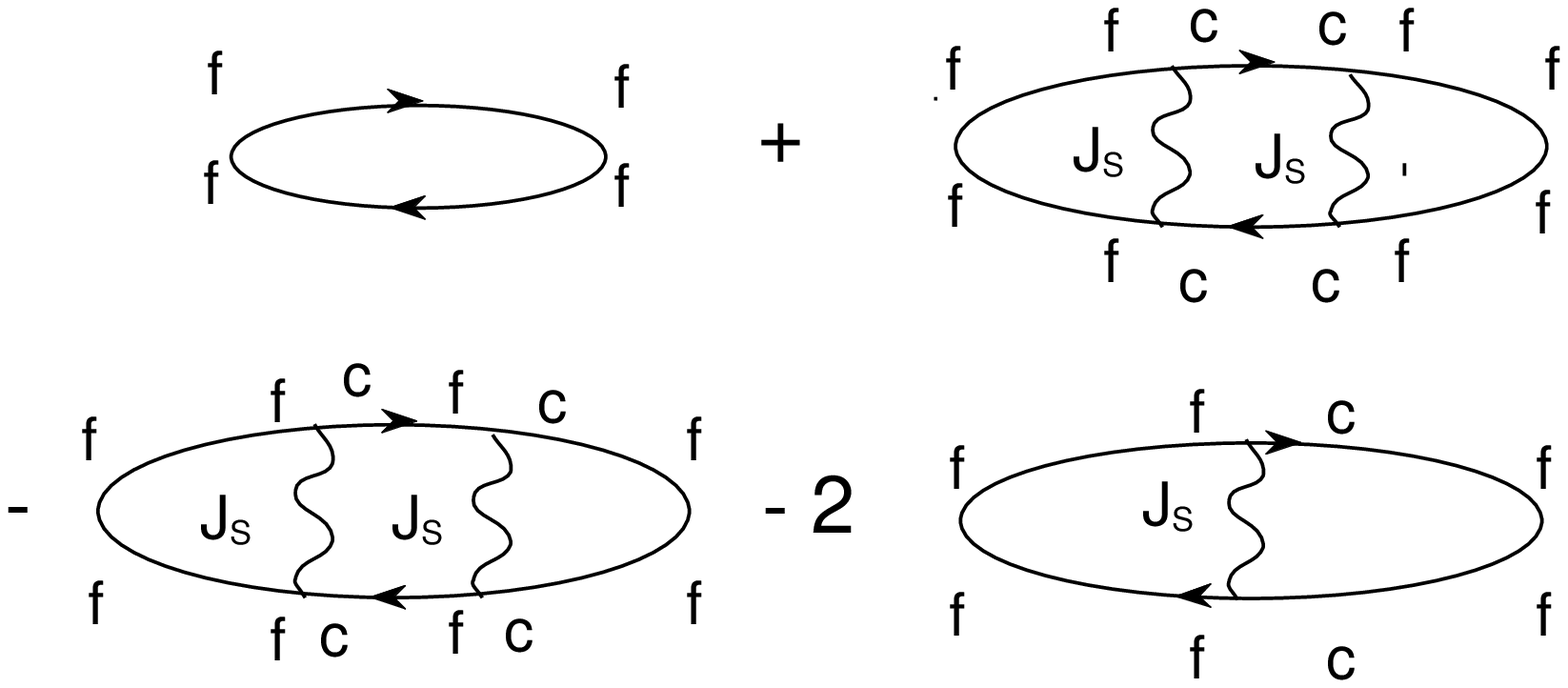,height=9cm,width=16cm}}
\vspace{2in}
\caption{}
\label{fig2}
\end{figure}

\vfill\eject

\vspace{2in}
\begin{figure}
\centerline{\psfig{file=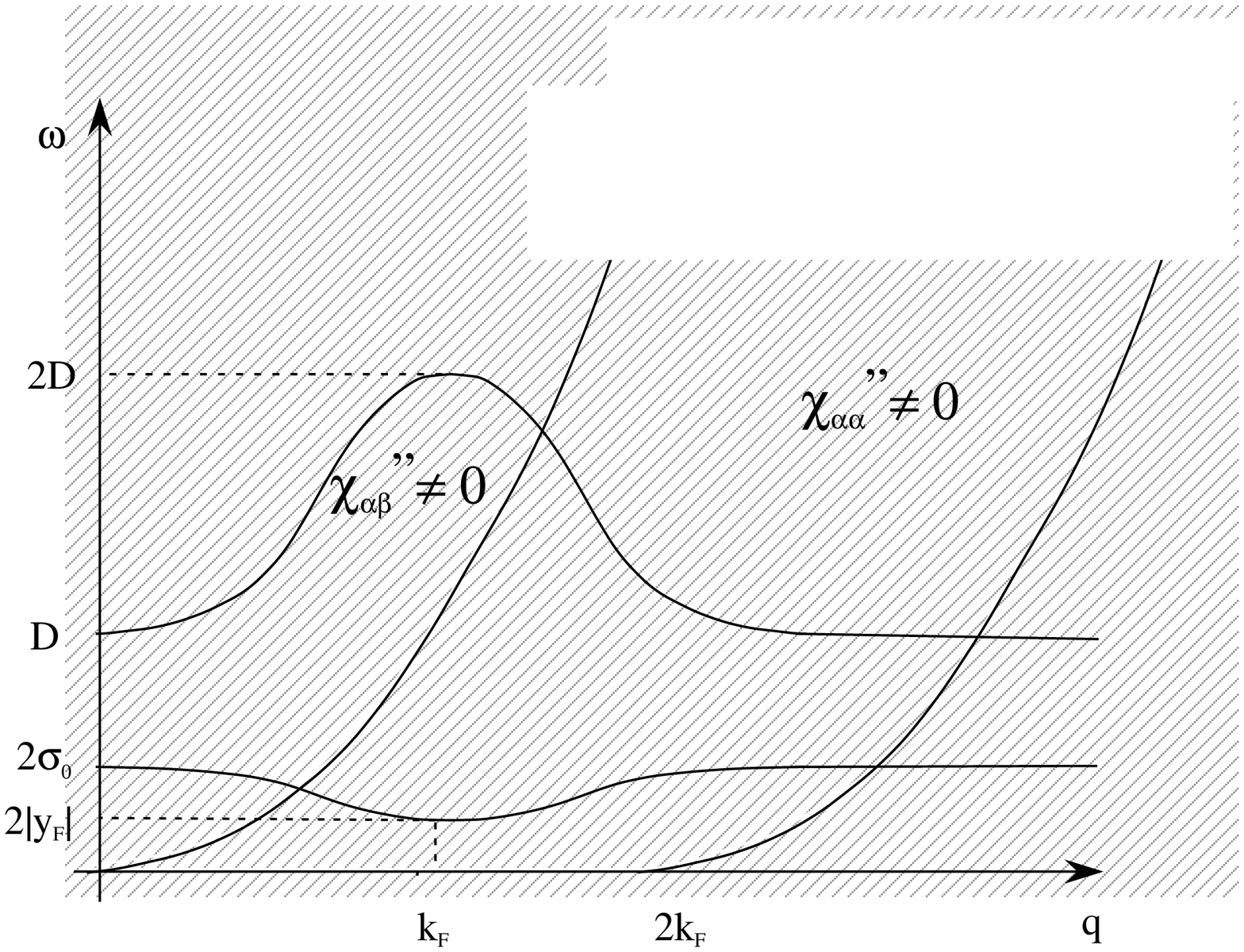,height=9cm,width=16cm}}
\vspace{2in}
\caption{}
\label{fig3}
\end{figure}

\vfill\eject

\end{document}